\documentclass[preprint,11pt]{elsarticle}
\graphicspath{{./Figs}}

\usepackage{amssymb}
\usepackage{nomencl}
\usepackage{etoolbox}
\usepackage{url}
\renewcommand\nomgroup[1]{%
  \item[\bfseries
  \ifstrequal{#1}{\vb{D}}{Subscripts}{%
  \ifstrequal{#1}{U}{Superscripts}{%
  \ifstrequal{#1}{S}{}{}}}%
]}
\newcommand{\epsV}{\varepsilon_V}
\newcommand{\perSol}{\CB{\vb{x}_0}}

\newcommand{\parameters}{\vb{\CB{a}}}
\newcommand{\hyperpar}{\boldsymbol{\CB{\alpha}}}
\usepackage[utf8]{inputenc}
\usepackage[margin=1.2in]{geometry}
\usepackage{amssymb}
\usepackage{nomencl}
\usepackage{etoolbox}
\usepackage{siunitx}
\let\svqty\qty
\usepackage{physics}
\let\qty\svqty
\usepackage{amsmath}
\usepackage{amssymb}
\usepackage{xcolor}
\usepackage[overload]{empheq}
\usepackage{fancyvrb}
\usepackage{cancel}
\usepackage{caption}
\usepackage{subcaption}
\usepackage{multirow}
\usepackage{multicol}
\usepackage{algorithm}
\usepackage{algpseudocode}
\usepackage{afterpage}
\usepackage{lipsum}
\usepackage{pdfpages}
\usepackage{accents}

\newcommand{\CB}[1]{{\color{black}{#1}}}

\newcommand{\map}{\mathrm{MP}}

\newcommand*{\colorboxed}{}
\def\colorboxed#1#{%
  \colorboxedAux{#1}%
}
\newcommand*{\colorboxedAux}[3]{%
  \begingroup
    \colorlet{cb@saved}{.}%
    \color#1{#2}%
    \boxed{%
      \color{cb@saved}%
      #3%
    }%
  \endgroup
}

\usepackage{tikz}

\newcommand{\Yoko}[1]{{\color{black}{#1}}}

\usepackage{lineno}
\journal{arXiv}
\begin{document}

\begin{frontmatter}



\title{Bayesian inference of physics-based models of acoustically-forced laminar premixed conical flames}


\author[a,b]{Alessandro Giannotta}
\author[b]{Matthew Yoko}
\author[a]{Stefania Cherubini}
\author[a]{Pietro De Palma}
\author[b]{Matthew P. Juniper\corref{corr1}}
\cortext[corr1]{Corresponding author.}
\ead{mpj1001@cam.ac.uk}
\affiliation[a]{organization={Department of Mechanics, Mathematics and Management, Politecnico di Bari},
            addressline={via Re David, 200},
            city={Bari},
            postcode={70125},
            country={Italy}}

\affiliation[b]{organization={Department of Engineering, University of Cambridge},
            addressline={Trumpington Street},
            city={Cambridge},
            postcode={CB2 1PZ},
            country={UK}}

\begin{abstract}
We perform twenty experiments on an acoustically-forced laminar premixed Bunsen flame and assimilate high-speed footage of the natural emission into a physics-based model containing seven parameters. 
The experimental rig is a ducted Bunsen flame supplied by a mixture of methane and ethylene. 
A high-speed camera captures the natural emission of the flame, from which we extract the position of the flame front.
We use Bayesian inference to combine this experimental data with our prior knowledge of this flame's behaviour. 
This prior knowledge is expressed through (i) a model of the kinematics of a flame front moving through a model of the perturbed velocity field, and (ii) \emph{a priori} estimates of the parameters of the above model with quantified uncertainties. 
%
%
%
We find the most probable \textit{a posteriori} model parameters using Bayesian parameter inference, and quantify their uncertainties using Laplace's method combined with first-order adjoint methods. This is substantially cheaper than other common Bayesian inference frameworks, such as Markov Chain Monte Carlo. 
This process results in a quantitatively-accurate physics-based reduced-order model of the acoustically forced Bunsen flame for injection velocities ranging from  $1.75\,\mbox{m/s}$ to $2.99 \,\mbox{m/s}$ and equivalence ratio values ranging from $1.26$ to $1.47$, using seven parameters.
We use this model to evaluate the heat release rate between experimental snapshots, to extrapolate to different experimental conditions, and to calculate the flame transfer function and its uncertainty for all the flames.
Since the proposed model relies on only seven parameters, it can be trained with little data and successfully extrapolates beyond the training dataset. Matlab code is provided so that the reader can apply it to assimilate further flame images into the model.
%
%


\end{abstract}




\end{frontmatter}
\section*{Novelty and Significance Statement}
Thermoacoustic systems tend to be extremely sensitive to small parameter changes, which makes them difficult to model \textit{a priori} from existing models in the literature. This means, however, that thermoacoustic models tend to be easy to train using data-driven methods because, with well-chosen experiments, their parameters can be easily observed from experimental data. 
This paper presents a novel use of Bayesian inference to combine experimental measurements, numerical simulations, and prior knowledge about flame behaviour. We outline our methodology and demonstrate its effectiveness using a laminar premixed Bunsen flame. 
%
%
Our approach yields a quantitatively-accurate physics-based model that predicts the expected value and uncertainty bounds of the flame transfer function between velocity and heat release rate perturbations. 
The proposed model contains only seven physical parameters, which is fewer parameters than non-physics-based models, and can therefore be trained on relatively little data.
We also illustrate how the trained model effectively extrapolates beyond the training dataset.
Our numerical code and experimental data are open access. 

\section{Introduction}
\label{sec:Introduction}


The efficiency of the mechanism driving thermoacoustic oscillations depends strongly on the phase difference between the heat release rate (h.r.r) and pressure oscillations \cite{Rayleigh1878}.
In turn, this phase difference usually depends strongly on the flame's dynamics \cite{Lieuwen2005} and on the acoustics of a combustion chamber, particularly if time delays in the flame are similar to or greater than the period of acoustic chamber modes \cite{Juniper2018}.
Many factors affect the flame dynamics and combustion chamber acoustics, meaning that a combustor's thermoacoustic behaviour tends to be extremely sensitive to small changes \cite{mongia2003challenges}.
For the same reason, the outputs of faithful thermoacoustic models are sensitive to small changes in models and their parameters.
On the negative side, this makes thermoacoustic systems difficult to model \emph{a priori} from existing models in the literature. 
On the positive side, this makes thermoacoustic models easy to train from experimental data because their parameters tend to be easily observable from experimental data.
With well-chosen experiments, we can therefore (i) tune the parameters of candidate models and (ii) compare candidate models against each other and select the one with most evidence, given the experimental data \cite{juniper2022JSV,Yoko2023}.
Also on the positive side, this extreme sensitivity means that thermoacoustic systems can often be stabilized by making small changes, which is attractive in industrial settings.
The challenge is to model systems accurately and to determine stabilizing modifications as quickly and accurately as possible.

The most influential component of a thermoacoustic system is the mechanism linking acoustic velocity and pressure fluctuations at the base of the flame to subsequent h.r.r. fluctuations in the body of the flame.
(In most settings, only the acoustic velocity fluctuations are influential.)
This mechanism is difficult to model or simulate \cite{poinsot2017prediction}, so researchers often rely on experimental measurements of the h.r.r. as a function of the acoustic velocity.
The fluctuating natural emission from the flame is not, however, a reliable measurement of the fluctuating h.r.r.\cite{han2015spatial}.
Alternatives, such as PLIF to identify reaction zones \cite{yuan2015reaction} are more accurate but are technically difficult and, in large systems, impractical.
This paper seeks to circumvent the above problems by combining experimental measurements with numerical simulations of a flame model. 
The flame model is physics-based and qualitatively-accurate. It gives, amongst other things, the flame front dynamics under harmonic forcing.
\CB{
The physics-based model has several parameters, whose expected values and uncertainties are inferred from the flame position, as found from experimental natural emission measurements.
We then use the model to calculate the h.r.r., and its uncertainty, as a function of the acoustic velocity perturbations. }

In previous work, Juniper \& Yoko \cite{juniper2022JSV} applied this process to a hot wire Rijke tube. 
From several candidate models in the literature, they selected those with the most evidence, given the data, and created a quantitatively accurate physics-based model of this thermoacoustic system.
In subsequent work on the same system, they used Bayesian optimal experimental design \cite{Yoko2024a} to evaluate the most informative experiments.
Following this, they extended the method to infer the flame response of laminar flames from pressure measurements \cite{Yoko2024}. This study revealed that, in some cases, the uncertainty in the inferred flame responses could be large when the source of information was pressure measurements alone.  
In this paper we revisit the flames studied in \cite{Yoko2024} in order to assess an alternative source of information about the flame response: the dynamics of the flame front position.
We use a high-speed camera to record the natural emission of the flames in steady conditions and then during a forced limit cycle. 
We assimilate this data into a physics-based model of a conical Bunsen flame from a previous study \cite{GiannottaCnF2023}. 
This model contains sufficient parameters to describe the forced experimental conical flame and is differentiable with respect to its parameters. 
We can therefore obtain the first derivatives of the model outputs with respect to model parameters and use these gradients to (i) find the most probable parameter values, given some data, and (ii) quantify our uncertainty in those parameters and the resulting uncertainty in the model predictions. 
%
\Yoko{This process creates a quantitatively-accurate physics-based model of the h.r.r. fluctuations as a function of the velocity perturbation. Unlike the output of many other data-driven approaches, the resulting model is interpretable, trustworthy and extrapolatable.}
%





The paper is structured as follows: in section \CB{\ref{sec:Experiments}} we explain the experiments and the image processing; in section \CB{\ref{sec:Model}} we report the key feature of the model and how we generate the gradients of the model output with respect to its parameters; in section \CB{\ref{sec:BayesianDataAssimilation}} we explain the Bayesian inference methodology and Laplace's approximation; in section \CB{\ref{sec:STEP1} and \ref{sec:STEP2} we show how we assimilate the experimental data to create the quantitatively-accurate physics-based model of the flame}; in section \CB{\ref{sec:FTF}}  we show how we employ the model to predict the flame transfer function and its uncertainty; in section \CB{\ref{sec:Extrapolation}} we show how the model is capable of extrapolating outside the training dataset; in section \CB{\ref{sec:Conclusions}} we discuss the conclusions and outline some future work.

\section{Experiments}\label{sec:Experiments}
The experimental observations consist of high-speed footage of a conical flame under steady and acoustically-forced conditions. The footage is processed to isolate the position of the flame front.

\subsection{Data Collection}

\begin{figure}
\setlength{\fboxsep}{0pt}%
\setlength{\fboxrule}{0pt}%
\begin{center}
\includegraphics[width = 0.75\textwidth]{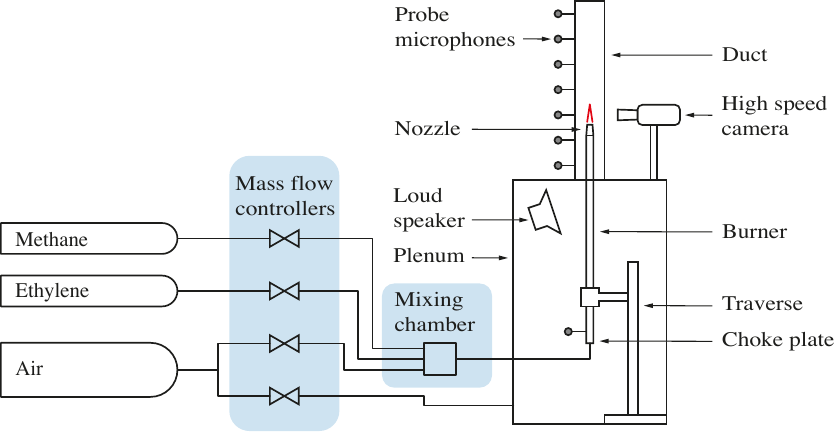}
\end{center}
\caption{Diagram of the experimental rig.}\label{fig:Experiment}
\end{figure}

The experimental configuration is a laminar premixed conical flame inserted into a vertical duct, as illustrated in figure \ref{fig:Experiment}. The lower end of the duct is fixed to a plenum chamber, through which co-flow air is supplied. The upper end is open to the atmosphere. The duct is a 0.8 \si{\meter} long section of quartz tube with an internal diameter of 75 \si{\milli\meter}. Quartz is used to allow for optical access to the flame.

The burner is a 0.85 \si{\meter} long section of brass tubing with an internal diameter of 14 \si{\milli\meter}. The outlet of the burner is fitted with a nozzle that is chosen such that the system could become thermoacoustically unstable. At the injection plane, the nozzle diameter is 9.35 \si{\milli\meter}. The burner is fuelled by a mixture of methane and ethylene over a wide range of equivalence ratios and fuel mass flow rates, which allow us to study flames with a wide range of thermoacoustic responses. The reactants are metered using a set of mass flow controllers, from which they flow through a choke plate, decoupling the supply lines from the acoustic fluctuations in the rig.


We use a high-speed camera to record the flame under both steady and perturbed conditions. If the system is linearly stable, we achieve the perturbed condition by using a loudspeaker to harmonically force the flame near the first resonant frequency of the system, $f = 230 \mbox{Hz}$. If the system is self-excited, we achieve the steady condition using active control with a phase-shift amplifier.

\begin{figure}
\centering
\includegraphics[width = 0.55 \textwidth]{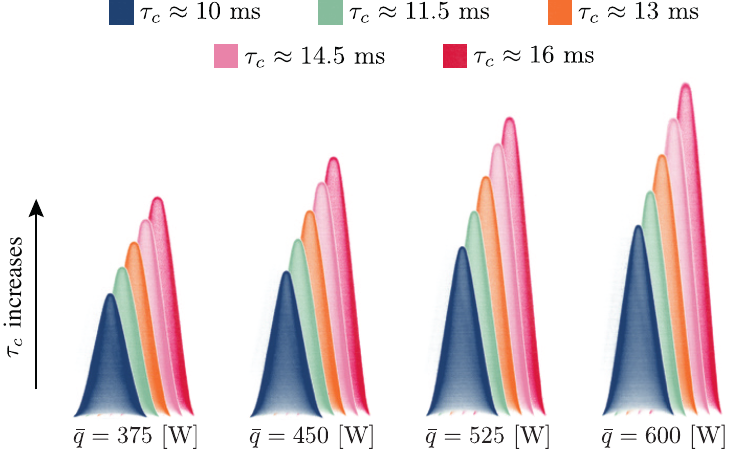}
\caption{Processed steady flame images from the 20 flames. Images are artificially coloured according to the approximate convective time delay. Flames with equal mean heat release rates are overlaid.}
\label{fig:flameLibraryIllustration}
\end{figure}

We study 20 flames, which we select to produce a wide range of flame responses. We parameterize the flames based on (i) the convective time delay, $\tau_c = L_f /\bar{u}$, which is the time taken for a perturbation travelling at the bulk velocity in the burner tube, $\bar{u}$, to traverse the length of the flame, $L_f$, and (ii) the mean heat release rate of the inner cone, $\bar{q}$. We group the 20 flames into five sets of four flames each, and select the composition of each flame such that the time delay is constant within a group and the mean heat release rate varies. The convective time delays range from 9.5 ms to 16 ms, and the mean heat release rates range from 375 W to 600 W. These flames produce a thermoacoustic behaviour ranging from strongly damped, to neutral, to strongly driven. 
%
The flame properties are summarized in table \ref{tab:flame_table} in \ref{sec:app_a}, and illustrated in figure \ref{fig:flameLibraryIllustration}.



\subsection{Data Processing}\label{sec:DataProcessing}
The raw footage is processed in order to (i) isolate the flame front position and (ii) produce a Euclidian distance field with the flame front as the zeroth level set.
The images are first averaged to improve the signal-to-noise ratio. For the steady flames, this involves averaging over 200 frames. For the unsteady flames, it involves phase-averaging over 20 frames captured at 10 phase angles.
The averaged images are straightened and centred so that the symmetry axis of the flame is vertical and in the centre of the frame. This is necessary to undo any misalignment in the camera setup. To do this, we exploit the symmetry of the flame and find the rigid transformation that maps the image to its mirror. From this, we can determine the transformation that moves the symmetry axis of the flame to be vertical and centred in the frame.
Once the flame is centred, we perform an Abel deconvolution to undo the line-of-sight integration of the natural luminosity of the flame. We use an implementation of the `onion-peeling' deconvolution \cite{Dasch1992}. This process produces a clearly defined flame front but accumulates noise towards the centre line of the image.
We perform image segmentation on the Abel deconvolved image using Chan-Vese segmentation \cite{Chan1999}. This separates the image signal from the background and allows us to interrogate the connectivity map of each cluster of pixels. We discard weakly connected clusters to denoise the image, leaving just the flame front as the dominant image structure. 
Finally, we perform a Euclidian distance transform on the binary image of the isolated flame front. This produces a matrix of the same dimensions as the original image, where each element contains the normal distance between the corresponding pixel and the flame front. We use this in our data assimilation process to quantify the discrepancy between the predicted and observed flame front positions.

\section{Reduced-Order Model of a laminar premixed Bunsen flame} \label{sec:Model}
\subsection{Model description}
The flame model is detailed in Ref. \citep{GiannottaCnF2023} and can be expressed compactly through:
\begin{equation} \label{eq:compactGovEq}
\dot{\vb{x}}(t) = \vb{f}(\vb{x}(t),t;\parameters), \quad \CB{\vb{x}(0) = \vb{x}_0},
\end{equation}
where $\vb{x}(t) = \left(\vb{r}(t),\vb{z}(t)\right)$ is the state vector defining the flame front position in terms of radial coordinates, $\vb{r}$, and longitudinal coordinates, $\vb{z}$, \Yoko{at a given time, $t$}, and \CB{$\vb{x_0}$ is the initial state.}
The vector $\parameters$ contains the physics-based parameters, while $\vb{f}$ is the physics-based nonlinear operator encapsulating the flame front dynamics. 
The main feature of the model is that it has been designed to be differentiable with respect to the state and the parameters.
The heat release rate $q(t)$ can also be expressed in compact form through:
\begin{equation} \label{eq:compacthrr}
q(t) = g(\vb{x}(t);\parameters),
\end{equation}
where $g$ is a nonlinear operator that links the heat release rate $q(t)$ to the flame front position and shape defined by $\vb{x}(t)$.
The model parameters $\parameters$ are: 
(i) the flame aspect ratio $\beta = \beta(\bar{u},S_L)$, which depends on the bulk velocity in the burner tube $\bar{u}$ and the unstretched laminar flame speed $S_L$; 
(ii) the nondimensional Markstein length $\mathcal{M}$, which depends on the thermal expansion parameter, the effective Lewis number of the mixture, the Zel'dovich number and the thermal conductivity of the mixture \citep{matalon2007intrinsic}; 
(iii) the shape parameter $\mu$, which linearly combines a uniform and a parabolic mean velocity profile \citep{yu2021data}; 
(iv) the wavelength of the harmonic perturbation velocity field $K$; 
(v) the amplitude of the acoustic forcing $\epsV$; 
(vi) the amplitude of the flame base oscillations $\lambda$; 
(vii) the initial phase of the flame base oscillations $\phi_0$.
These seven parameters are sufficient to describe the flame front dynamics qualitatively. 
All lengths, including the Markstein length, are normalised by the nozzle radius $R = 4.675$ \si{\milli\meter}.

\subsection{Obtaining periodic solutions and their sensitivities using adjoint methods}
\CB{
All flames are perturbed by a harmonic acoustic forcing with period $T$. 
%
%
We consider the following vector-valued function:
\begin{equation} \label{eq:periodicEq}
\vb{F}(\perSol;\parameters) = \vb{x}(T) - \perSol, \quad \text{where } \vb{x}(T) = \int_{0}^{T} \vb{f}(t,\vb{x};\parameters)\, \mbox{d}t + \perSol.
\end{equation}
For a given parameter set $\parameters$, a solution $\vb{x}(t)$ of equation \eqref{eq:compactGovEq} with an initial state $\perSol$ is periodic with period $T$ if:
\begin{equation} \label{eq:periodic2}
\vb{F}(\perSol;\parameters) =  \vb{0}.    
\end{equation}
We solve equation \eqref{eq:periodic2} using the trust-region dogleg algorithm and by providing the partial derivative of $\vb{F}$ with respect to the initial state $\perSol$
\begin{equation}
\pdv{\vb{F}}{\perSol} = \pdv{\vb{x}(T)}{\perSol} - \vb{I},
\end{equation}
where $\vb{I}$ is the identity matrix, and $\partial{\vb{x}(T)}/\partial{\perSol}$ is the partial derivative of the state at time $T$ with respect to changes in the initial state $\perSol$, which we compute using adjoint methods.
We can also obtain the gradient of the periodic solution $\CB{\vb{x}(t)}$ with respect to the parameters $\parameters$, by differentiating equation \eqref{eq:periodicEq} with respect to $\parameters$:
\begin{equation}
\begin{split}
\dv{\vb{F}}{\parameters} = &\pdv{\vb{x}(T)}{\parameters} + \pdv{\vb{x}(T)}{\perSol}\dv{\perSol}{\parameters} - \dv{\perSol}{\parameters} = \vb{0}\\
\implies &\dv{\perSol}{\parameters} = -\left(\pdv{\vb{x}(T)}{\perSol} - \vb{I}\right)^{-1}\pdv{\vb{x}(T)}{\parameters},
\end{split}
\end{equation}
and then by differentiating:
\begin{equation}
 \dv{\vb{x}(t)}{\parameters} = \pdv{\vb{x}(t)}{\parameters} + \pdv{\vb{x}(t)}{\perSol} \dv{\perSol}{\parameters}.
\end{equation}
Therefore we compute the gradients of the h.r.r. with respect to the parameters by differentiating \eqref{eq:compacthrr}:
\begin{equation}
 \dv{q(t)}{\parameters} = \pdv{g(\vb{x}(t);\parameters)}{\parameters} + \pdv{g(\vb{x}(t);\parameters)}{\vb{x}(t)} \dv{\vb{x}(t)}{\parameters}.
\end{equation}
This quantity is essential for quick calculation of the uncertainties of the heat release rate and therefore the flame transfer function. 
}
\section{Bayesian data assimilation}\label{sec:BayesianDataAssimilation}

Bayesian data assimilation provides us with a set of tools that allow us to (i) infer the most probable parameters of a physics-based model, given some data, (ii) quantify the uncertainty in the model parameters, and estimate the systematic uncertainty in the model and data, (iii) rank several candidate models to select the best model, given the data, and (iv) identify optimal experiment designs and sensor placements. In this paper, we use points (i) and (ii), so these are described in detail. A demonstration of model ranking can be found in \cite{Yoko2024}, and a demonstration of optimal experiment design can be found in \cite{Yoko2024a}.

\CB{
In this section, we introduce the following notation to outline the Bayesian inference methodology: 
}
the data $\vb{D}$ contains experimental observations of the flame position in steady and/or unsteady configurations; 
the model $\mathcal{H}$ encodes a physics-based reduced-order model, such that for a given set of parameters, $\vb{a}$, the model $\mathcal{H}(\vb{a})$ gives a prediction of the data $\vb{D}$. 
%

\subsection{Parameter inference}
When performing parameter inference, we assume that the model is structurally correct and we infer its most probable parameters, $\vb{a}_\map$. The model is rarely free of structural error, however, and we will revisit this assumption later. We encode our level of uncertainty in the parameter values through a probability distribution, which we denote $P(\bullet)$. Using any prior knowledge we have about the unknown parameters (which may be none at all), we propose a prior probability distribution over the parameter values, $P(\mathbf{a}|\mathcal{H})$. We then assimilate the data, $\vb{D}$, by performing a Bayesian update on the parameter values:
\begin{equation}\label{eq:BayesRule}
P(\vb{a}|\vb{D},\mathcal{H}) = \frac{P(\vb{D}|\vb{a},\mathcal{H})P(\vb{a}|\mathcal{H})}{P(\vb{D}|\mathcal{H})}
\end{equation}
The quantity on the left-hand side of equation~\eqref{eq:BayesRule} is the {posterior} probability of the parameters, given the data. {It is generally computationally intractable to calculate the full posterior, because this requires integration over a large parameter space. The integral typically cannot be evaluated analytically, and requires thousands of model evaluations to compute numerically. At the parameter inference stage, however, we are only interested in finding the most probable parameters, which are those that maximize the posterior. We therefore use an optimization algorithm to find the peak of the posterior without evaluating the full distribution. This process is made computationally efficient by (i)~assuming that the experimental uncertainty is Gaussian distributed, and (ii)~choosing the prior parameter distribution to be Gaussian. Assumption (i) is reasonable for well-designed experiments in which the uncertainty is dominated by random error, which is typically Gaussian distributed. For assumption (ii) we note that the choice of prior is often the prerogative of the researcher, and we are free to exploit the mathematical convenience offered by the Gaussian distribution.} {The correlations between model parameters are rarely known \textit{a-priori}, so an independent Gaussian distribution is often used for the prior, as in this paper.}
When finding the most probable parameters, we neglect the denominator of the right-hand side of equation~\eqref{eq:BayesRule}, because it does not depend on the parameters. It is then convenient to define a cost function, $\mathcal{J}$, to be the negative log of the numerator of equation~\eqref{eq:BayesRule}, which we minimize. 
\CB{
In this study, the data, $\vb{D}$, consists of a set of experimental observations of the flame position at various time frames and experimental configurations.
%
In its most general form, the cost function \Yoko{for a single image of a flame} is:
\begin{equation}\label{eq:CostFunc_general}
\begin{split}
\mathcal{J}(\vb{a}) =& -\log{\{P(\vb{D}|\vb{a},\mathcal{H})P(\vb{a}|\mathcal{H})}\} \\
  =& \frac{1}{2} (\vb{s}(\vb{a})-\vb{z})^T \vb{C}_{ee}^{-1} (\vb{s}(\vb{a})-\vb{z})\\
&+\frac{1}{2} (\vb{\vb{a}-\vb{a_p}})^T \vb{C}_{aa}^{-1} (\vb{\vb{a}-\vb{a_p}}) + \text{const.}
\end{split}
\end{equation}

%
\Yoko{where: $\vb{s}(\vb{a})$ and $\vb{z}$ are column vectors of the predicted and measured flame front positions respectively; $\vb{C}_{ee}$ is the covariance matrix describing the experimental uncertainty; $\vb{a}$ and $\vb{a}_p$ are column vectors of the current and prior parameter values respectively; $\vb{C}_{aa}$ is the covariance matrix describing the uncertainty in the prior, and const. is a constant that arises from the Gaussian pre-exponential factors, which does not impact the maximum posterior parameter estimation $\vb{a}_\map$. 
}
%
%
In this study, we evaluate the discrepancy between the model predictions and the experimental observations by projecting the predicted flame front position onto the Euclidian distance field produced from the corresponding frame of the high speed footage. This produces a column vector in which each entry is the normal distance between the predicted and observed flame front position at each point in the model.}

We see from equation~\eqref{eq:CostFunc_general} that assuming Gaussian distributions for the data and prior reduces the task of parameter inference to a quadratic optimization problem. \CB{We solve this optimization problem using the BFGS algorithm, where the gradient information is provided using adjoint methods}. 
%

\begin{figure}[!htp]
    \centering
    \includegraphics[width=\textwidth]{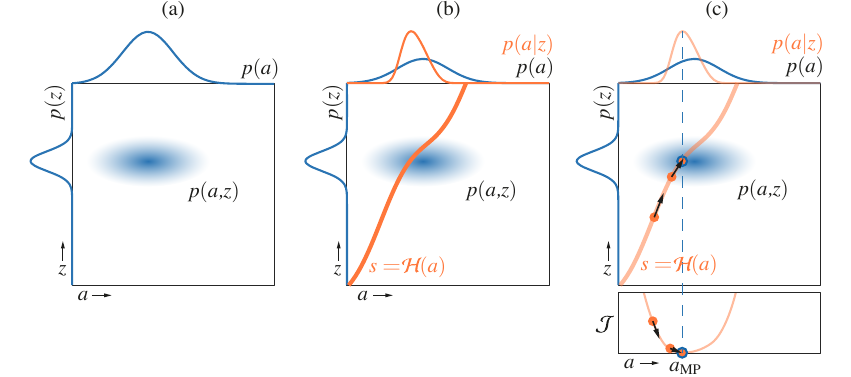}
    \caption{{Illustration of parameter inference on a simple univariate system. (a) the
    marginal probability distributions of the prior and data, $p(a)$ and $p(z)$, as well as their
    joint distribution, $p(a,z)$ are plotted on axes of parameter value, $a$, vs observation
    outcome, $z$. (b) the model, $\mathcal{H}$, imposes a functional relationship between the parameters,
    $a$, and the predictions, $s$. Marginalizing along the model predictions yields the true
    posterior, $p(a|z)$. This is computationally intractable for even
    moderately large parameter spaces. (c) instead of evaluating the full posterior, we use
    gradient-based optimization to find its peak. This yields the most probable parameters,
    $a_\map$}}
    \label{fig:ParameterInference}
\end{figure}

{The parameter inference process is illustrated in figure~\ref{fig:ParameterInference} for a simple system with a single parameter, $a$, and a single observable variable, $z$. In (a) we show the marginal probability distributions of the prior, $p(a)$, and the data, $p(z)$. The prior and data are independent, so we construct the joint distribution, $p(a,z)$ by multiplying the two marginals. In (b), we overlay the model predictions, $s$, for various values of $a$. Marginalizing along the model predictions yields the true posterior, $p(a|z)$. This is possible for a cheap model with a single parameter, but exact marginalization quickly becomes intractable as the number of parameters increases. In (c) we plot the cost function, $\mathcal{J}$, which is the negative log of the unnormalized posterior. We show the three steps of gradient-based optimization that are required to find the local minimum, which corresponds to the most probable parameters, $a_\mathrm{MP}$.}

\subsection{Uncertainty quantification}
Uncertainty quantification can be split into two steps: (i) quantifying the parametric uncertainty and propagating it to the model prediction uncertainty, and (ii) estimating the systematic and structural uncertainty in the experiments and model predictions. We will deal with these separately.

\subsubsection{Parameteric uncertainty}
Once we have found the most probable parameters, $\vb{a}_\map$, we estimate the uncertainty in these parameter values using Laplace's method \cite{Jeffreys1973,mackay1992information,juniper2022JSV}. This method approximates the posterior as a Gaussian distribution with a mean of $\mathbf{a_\map}$, and a covariance given by the Hessian of the cost function:
\begin{equation}
\begin{split}
  \vb{C}_{aa}^{\map-1} &\approx \frac{\partial^2\mathcal{J}}{\partial a_i \partial a_j}
\end{split}
\end{equation}

The Hessian of the cost function is estimated by the BFGS algorithm during the minimization process. This estimate is based on the change in the value and gradient of the cost function over the previous iteration. 

\begin{figure}[!htp]
    \centering
    \includegraphics[width=\textwidth]{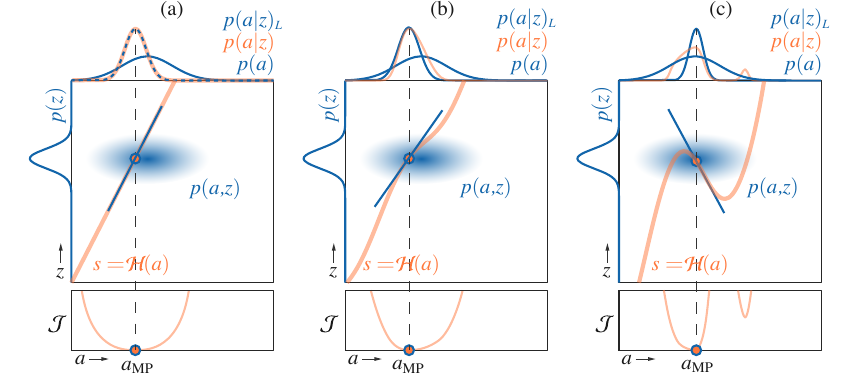}
    \caption{{Illustration of uncertainty quantification for three univariate systems, comparing the true posterior, $p(a|z)$ to the approximate posterior from Laplace's method, $p(a|z)_L$. (a) the
    model is linear in the parameters, so the true posterior is Gaussian and Laplace's method
    is exact. (b) the model is weakly nonlinear in the parameters, the true posterior is slightly
    skewed, but Laplace's method yields a reasonable approximation. (c) the model is strongly
    nonlinear in the parameters, the posterior is multi-modal and Laplace's method
    underestimates the uncertainty}}
    \label{fig:UncertaintyQuantification}
\end{figure}

{The accuracy of \Yoko{the Laplace} approximation depends on the functional dependence between the model predictions and the parameters. This is shown graphically in figure~\ref{fig:UncertaintyQuantification} for three univariate systems. In (a), the model is linear in the parameters. Marginalizing a Gaussian joint distribution along any intersecting line produces a Gaussian posterior, so Laplace's method is exact. In (b), the model is weakly nonlinear in the parameters. The true posterior is skewed, but the Gaussian approximation is still reasonable. This panel also shows a geometric interpretation of Laplace's method: the approximate posterior is given by linearizing the model around $\mathbf{a}_\mathrm{MP}$, and marginalizing the joint distribution along the linearized model. In (c), the model is strongly nonlinear in the parameters, so the true posterior is multi-modal and the main peak is highly skewed.} {In this case, the gradient-based optimization algorithm will only find a single local minimum, which will depend on the choice of initial condition for the optimization. Furthermore, we see that the posterior estimated using Laplace's method is a poor approximation of the true posterior. This problem can be avoided by reducing the extent of the nonlinearity captured by the joint distribution by (i) shrinking the joint distribution by providing more precise prior information or more precise experimental data, or (ii) re-parameterizing the model to reduce the strength of the nonlinearity \citep[Chapter 27]{mackay1992information}.}

{In many cases, the use of this approximate inference framework approximation will be justifiable, given the substantial reduction in computational cost compared to sampling methods, which are the only viable alternative for constructing the posterior. In previous work, we compared the computational cost of our framework to two sampling approaches \cite{Yoko2024}. The comparison was done on a computationally cheap thermoacoustic network model. Applying our framework to this model, we can compute the posterior probability of five unknown parameters in under 5 seconds on a single core on a laptop. The same inference problem takes 35 CPU hours running on a workstation when solved with Markov Chain Monte Carlo, and 22 CPU hours when solved with Hamiltonian Monte Carlo.}

\subsubsection{Uncertainty propagation}

To quantify the uncertainty in the model predictions due to the parametric uncertainty, we linearise the model around $\vb{a}_\map$ and propagate the parameter uncertainties through the linearised model \cite{Putko2002}. The uncertainty in the model prediction is given by:
\begin{equation}
\label{eq:modelUnc_1}
    \mathbf{C}_{xx} = \vb{J}^T\vb{C}_{aa}\vb{J}
\end{equation}
where $\mathbf{C}_{xx}$ represents the covariance in the model predictions, \CB{and $\vb{J}$ denotes the Jacobian matrix of the model predictions with respect to the parameters $\vb{a}$.} The marginal uncertainty in each model prediction, $(\sigma_{x_i})^2$, is given by the diagonal entries of $\mathbf{C}_{xx}$.

\subsubsection{Systematic uncertainty}

In most cases, experimental data will contain some systematic uncertainty, and models will contain some structural uncertainty. These uncertainty sources cannot be quantified \emph{a-priori}, and are often referred to as ``unknown unknowns''. We can, however, construct a total covariance matrix, {$\mathbf{C}_{tt}$, which encodes the total uncertainty due to (i)~the known experimental uncertainty, (ii)~the unknown systematic experimental uncertainty, and (iii)~the unknown structural model uncertainty. We can then estimate this total covariance from the posterior discrepancy between the model and the data. This must be done simultaneously with parameter inference, because the posterior parameter distribution depends on the total uncertainty in the model and data. We therefore replace $\mathbf{C}_{ee}$ with $\mathbf{C}_{tt}$ in equation~\eqref{eq:CostFunc_general}, and estimate the total uncertainty by simultaneously minimizing $\mathcal{J}$ with respect to $\mathbf{a}$ and $\mathbf{C}_{tt}^{-1}$. }

{We begin by calculating the derivative of $\mathcal{J}$ with respect to $\mathbf{C}_{tt}^{-1}$, assuming that the observed variables are uncorrelated, and keeping in mind that the normalizing constant, $K$, depends on $\mathbf{C}_{tt}$:}

\begin{equation}
    \begin{aligned}
        \mathcal{J} = \frac{1}{2}(\mathbf{s}(\mathbf{a})-\mathbf{z})^T \mathbf{C}_{tt}^{-1} (\mathbf{s}(\mathbf{a})-\mathbf{z}) &+ \log \left(\sqrt{(2\pi)^k|\mathbf{C}_{tt}|}\right)\\
        + \frac{1}{2}(\mathbf{a}-\mathbf{a_p})^T \mathbf{C}_{aa}^{-1} (\mathbf{a}-\mathbf{a_p}) &+ \log \left(\sqrt{(2\pi)^k|\mathbf{C}_{aa}|}\right)
    \end{aligned}
\end{equation}
\begin{equation}
    \label{eq:pJpC_ee}   
    {\frac{\partial \mathcal{J}}{\partial \mathbf{C}_{tt}^{-1}} = \frac{1}{2}(\mathbf{s}(\mathbf{a})-\mathbf{z})(\mathbf{s}(\mathbf{a})-\mathbf{z})^T \circ \mathbf{I}- \frac{1}{2}\mathbf{C}_{tt}}
\end{equation}
\noindent
{where $\mathbf{I}$ is the identity matrix, and $\circ$ denotes the Hadamard product. For a given set of parameters, the most probable $\mathbf{C}_{tt}$ sets equation~\eqref{eq:pJpC_ee} to zero. This gives the estimate:}

\begin{equation}
    {\mathbf{C}_{tt} = (\mathbf{s}(\mathbf{a})-\mathbf{z}) (\mathbf{s}(\mathbf{a})-\mathbf{z})^T \circ \mathbf{I}}
\end{equation}
\noindent
which is the expected result that the total variance in the model and data is the square of the discrepancy between the model predictions and the data. Although we cannot directly identify the source of the unknown uncertainty because the experimental and model uncertainties cannot be disentangled, the inferred total uncertainty can assist the researcher with identifying potential error sources. For example, if the unknown error in a single sensor is unexpectedly large, this could indicate a faulty sensor or bad installation. If the unknown error at a certain experimental operating condition is large, this could prompt the researcher to repeat that experiment. If the unknown error grows with one of the input variables, the researcher might investigate the model to see if any important physical phenomena has been neglected.

\section{\CB{Creating the quantitatively-accurate physics-based model}}\label{sec:Results}
\begin{figure}
\setlength{\fboxsep}{0pt}%
\setlength{\fboxrule}{0pt}%
\begin{center}
\includegraphics[width=0.8\textwidth]{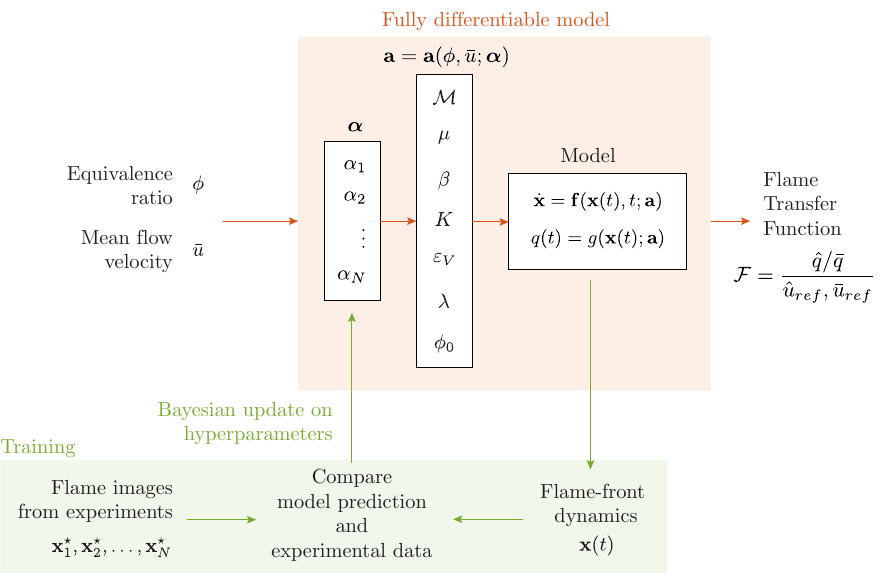}
\end{center}
\caption{Illustration showing the methodology for constructing a quantitatively-accurate reduced-order model.
The general model depends on the hyperparameters $\hyperpar$ and comprises two components: the physics-based reduced order model, which depends on the physical parameters $\parameters$, and a linking model $\parameters = \parameters(\phi,\bar{u};\hyperpar)$ that establishes the relationship between these physical parameters $\parameters$ and the hyperparameters $\hyperpar$. The model is fully differentiable and provides the flame front dynamics and the flame transfer function. By comparing the model's predictions of the flame front with experimental images, we refine our hyperparameters $\hyperpar$ through Bayesian updates. \Yoko{The resulting model predicts the flame transfer function, with confidence bounds, for any combination of equivalence ratio $\phi$ and the mean flow velocity $\bar{u}$.} This model is both interpretable and capable of extrapolation.}
\label{fig:Methodology_Diagram}
\end{figure}

\CB{
In this study, we perform Bayesian data assimilation in two steps. 
In the first step, we assimilate the flame front position data of each \CB{flame individually}. 
Therefore, for each experimental configuration (i.e. for each combination of mean flow velocity $\bar{u}$ and equivalence ratio $\phi$), we determine the most probable set of model parameters $\parameters = (\mathcal{M},\mu,\beta,K,\epsV,\lambda,\phi_0)$, which have been introduced in section \ref{sec:Model}. 
This process does not create a general model \Yoko{that can predict the behaviour of an arbitrary flame}, but is useful for assessing the influence of the model parameters $\parameters$ on the model predictions.
After completion of this step, we need to introduce a set of hyperparameters $\hyperpar$, such that $\parameters = \parameters(\phi,\bar{u};\hyperpar)$.
We then simultaneously assimilate the flame front position data of all flames.  This allows us to determine the most probable set of hyperparameters $\hyperpar$ so that, for each combination of mean flow velocity $\bar{u}$ and equivalence ratio $\phi$, the model $\parameters = \parameters(\phi,\bar{u};\hyperpar)$ provides the best prediction of the physics-based model parameters $\parameters$ and, consequently, of the flame front dynamics. 
In figure \ref{fig:Methodology_Diagram}, we show a diagram describing the methodology followed in this study. 
The end goal is to create a general model that provides the flame transfer function, along with quantified uncertainty bounds, for any given pair of equivalence ratio and bulk velocity in the burner tube.

}

\subsection{Step 1: assimilation of each \CB{flame} individually}\label{sec:STEP1}
First, we assimilate the steady-state snapshots in order to infer $\mathcal{M}$, $\beta$ and $\mu$. 
Second, we use these inferred values as priors for the second step, in which we assimilate the unsteady flame images and infer all parameters simultaneously: $\mathcal{M}$, $\beta$, $\mu$, $K$, $\varepsilon_V$, $\lambda$, $\phi_0$.
\CB{Here we show the process for flame 17 in Table \ref{tab:flame_table}}, which has an equivalence ratio $\phi = 1.34$, a flow rate of $\bar{u} = 2.86\, \mbox{m/s}$ and an equal volume fraction of methane $(\mathrm{C H_4})$ and ethylene $(\mathrm{C_2H_4})$.

\subsubsection{Steady-state data assimilation}
\begin{figure}
\setlength{\fboxsep}{0pt}%
\setlength{\fboxrule}{0pt}%
\begin{center}
\includegraphics[width = 0.85 \textwidth]{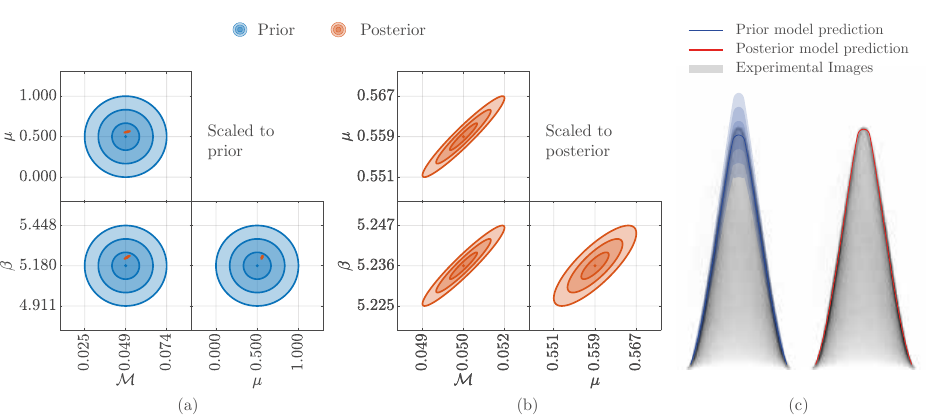}
\end{center}
\caption{(a-b) Probability distributions of parameters in the steady case: prior (blue) and posterior (red). Contours denote 1, 2, and 3 standard deviations from the prior $\parameters_{p}$ and posterior $\parameters_\map$ estimations. Parameters include the flame aspect ratio $\beta$, which is a function of the unstretched laminar flame speed $S_L$ and the bulk velocity in the burner tube $\bar{u}$, the shape parameter for the velocity field $\mu$, and the nondimensional Markstein length $\mathcal{M}$. Both panels represent the same quantities but panel (a) is scaled to the prior distributions, while panel (b) is scaled to the posterior distributions.
(c) Comparison between model predictions before (red line) and after (blue line) assimilating the experimental steady-state flame image. Confidence intervals, spanning 3 standard deviations, are illustrated by the red and blue shadings overlaying the steady flame image.}
\label{fig:SteadyPriorAndPosterior}
\end{figure}

%
The steady flame front position depends on $\beta$, $\mathcal{M}$ and $\mu$.
We can include prior knowledge about the parameter $\beta = \beta(\bar{u},S_L)$ because it depends on the unstretched flame speed $S_L$, which can be estimated using the open source chemical kinetics code Cantera \cite{goodwin2018cantera}, and the mean flow velocity $\bar{u}$, which can be measured during the experiments. Furthermore, we can estimate the value of the Markstein length by using Matalon's formulation \cite{matalon2007intrinsic}.
Following this process for flame $17$, we assign a prior estimate to the unstretched laminar flame speed of $S_L = 0.4036 \, \mbox{ms}^{-1}$ and a dimensional Markstein length of $\tilde{\mathcal{M}} =0.2579\, \mbox{mm}$, corresponding to $\mathcal{M} = \tilde{\mathcal{M}}/R = 0.0552$. \Yoko{To construct the parameter covariance matrix, we must assign confidence bounds to these values, which is naturally a subjective process. In this case, we assign an uncertainty of $ 6\sigma_{S_L} = 10\% S_L$ and $6\sigma_{\mathcal{M}} = 100\% \mathcal{M}$ to indicate that we are relatively more confident in our ability to predict laminar flame speed than Markstein length}.
The only prior knowledge available for the velocity shape parameter $\mu$ is that it can vary between $0$ and $1$, so we set a prior with large uncertainty for this parameter.
From the steady-state flame snapshot, we can infer the \textit{maximum a posteriori} probability estimate of the aspect ratio, $\beta$, the nondimensional Markstein length, $\mathcal{M}$, and the velocity shape parameter, $\mu$. Figure \ref{fig:SteadyPriorAndPosterior}(a-b) shows the prior and posterior probability distributions of the model parameters. The contours show $1,2$ and $3$ standard deviations from the prior $\parameters_{p}$ (blue) and posterior $\parameters_\map$ (red). We see that the experimental observations have reduced the uncertainties in the model parameters.
Figure \ref{fig:SteadyPriorAndPosterior}(c) shows the model prediction $\pm 3$ standard deviations overlaid on the steady flame image. The largest uncertainties are found at the flame tip, which is the region that is most sensitive to the parameters. The model prediction uncertainty is found by solving equation \eqref{eq:modelUnc_1}.


\subsubsection{Unsteady-state (acoustically forced)}
\begin{figure}
\setlength{\fboxsep}{0pt}%
\setlength{\fboxrule}{0pt}%
\begin{center}
\includegraphics[width=.85\textwidth]{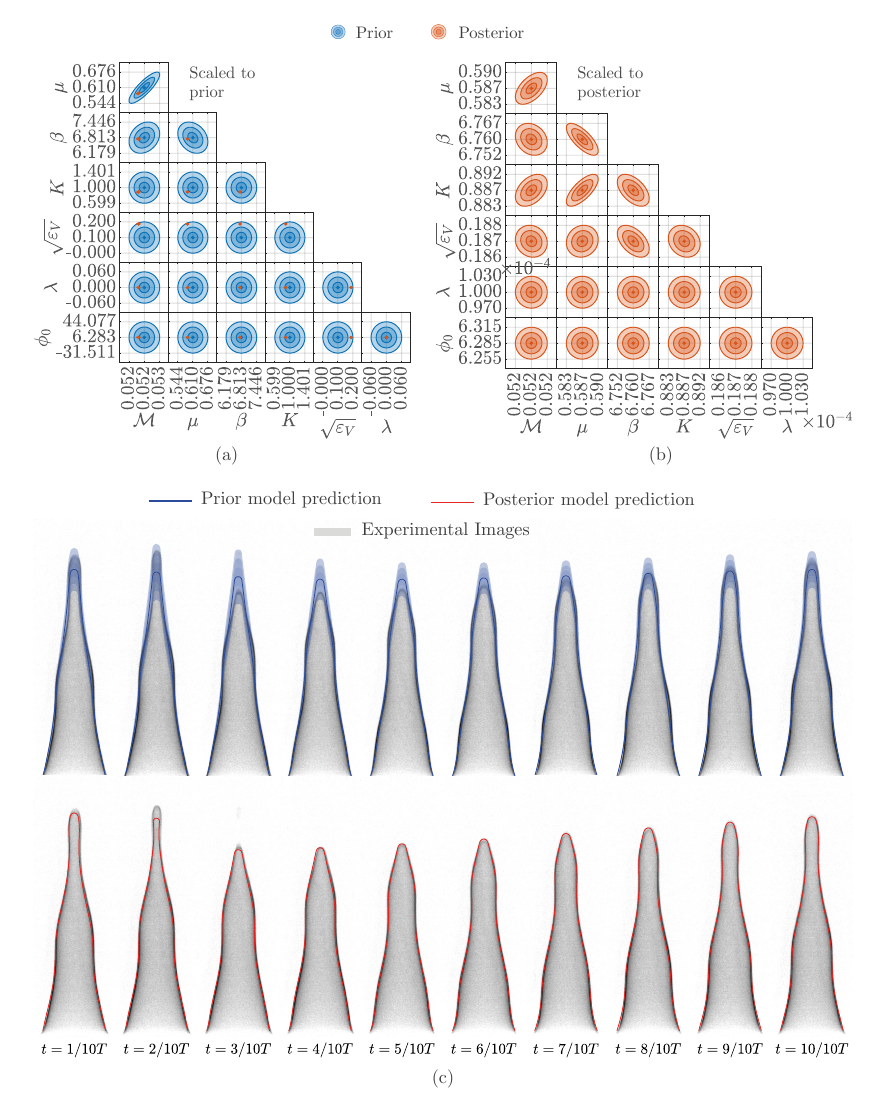}
\end{center}
\caption{(a-b) Probability distributions of parameters for the acoustically forced case: prior (blue) and posterior (red). Contours depict 1, 2, and 3 standard deviations from the prior $\parameters_{p}$ and posterior $\parameters_{\map}$ estimations. Both panels represent the same quantities; however, panel (a) is scaled to the prior distributions, while panel (b) is scaled to the posterior distributions. Parameters include the flame aspect ratio $\beta$,  which is a function of the unstretched laminar flame speed $S_L$ and the bulk velocity in the burner tube $\bar{u}$, the shape parameter for the velocity field $\mu$, the nondimensional Markstein length $\mathcal{M}$, the nondimensional velocity perturbation convective speed $K$, and the amplitude of velocity perturbations $\varepsilon_V$.
(c) Comparison of model predictions before (red line) and after (blue line) assimilating the experimental video footage. The confidence interval, spanning 3 standard deviations, is indicated by red and blue shadings overlaid on the experimental flame images.}
\label{fig:UnsteadyPriorAndPosterior}
\end{figure}

We repeat the process with the images of the acoustically forced flame. In this case, we set the values of the three steady parameters using the information gained from the previous step, and we set large uncertainties in the prior values of $K$, $\epsV$, $\lambda$ and $\phi_0$.
Figure \ref{fig:UnsteadyPriorAndPosterior}(a-b) shows the prior and posterior distribution of the parameters before and after assimilating the unsteady flame images.
The results show that the uncertainty in the values of $K$, $\epsV$, $\lambda$ and $\phi_0$ has been reduced significantly through the assimilation of this data, and the values of $\mathcal{M}$ and $\mu$ have shifted slightly from the values found during the assimilation of the steady data, without exceeding the 3-standard deviation bounds.
Figure \ref{fig:UnsteadyPriorAndPosterior}(c) shows the model prediction $\pm 3$ standard deviations, plotted against the unsteady flame images before (blue) and after (red) assimilating the experimental video footage. As in the steady-state case, the largest uncertainties are found at the flame tip. We see that the model matches the data well at each frame. With this, we have produced a digital twin of the flame, which we can use to estimate additional quantities of interest such as the unstretched laminar flame speed $S_L$, the Markstein length $\mathcal{M}$, the convective speed of velocity perturbations $K$ and the heat release rate, which were not directly measured in the experiments. Furthermore, because the model is probabilistic, we can also quantify our uncertainty in these estimates.

For the example of flame 17, the unstretched laminar flame speed and the Markstein length are estimated to be $S_L=0.4191 \,\mbox{m/s}$ and $\mathcal{M} = 0.2430 \,\mbox{mm}$ compared to the value of $S^c_L = 0.4036 \,\mbox{m/s}$ and $\mathcal{M}^c = 0.2579 \,\mbox{mm}$ obtained from Cantera and the value of the convective speed is estimated to be $K = 0.8874 $ compared to $K = 1$ in Schuller et al. \citep{schuller2002modelling}, $K = 0.9$ in  Kashinath et al. \citep{kashinath2014nonlinear} and $K = 0.83$ in Orchini \& Juniper \cite{Orchini2015}. In this case, we were able to provide reasonable prior information for $\mathcal{M}$ and $S_L$. Similar values are, however, inferred even if the prior is inaccurate and uncertain. This methodology could therefore be used to cheaply and easily estimate combustion properties of fuels for which data are not available. 




The proposed model contains few parameters, allowing it to be trained on a relatively small amount of data. Furthermore, the data required to train the model is easy to collect from a simple experiment because only snapshots of the natural emission of the flame are required.

\begin{figure}
\begin{center}
\includegraphics[width = .95\textwidth]{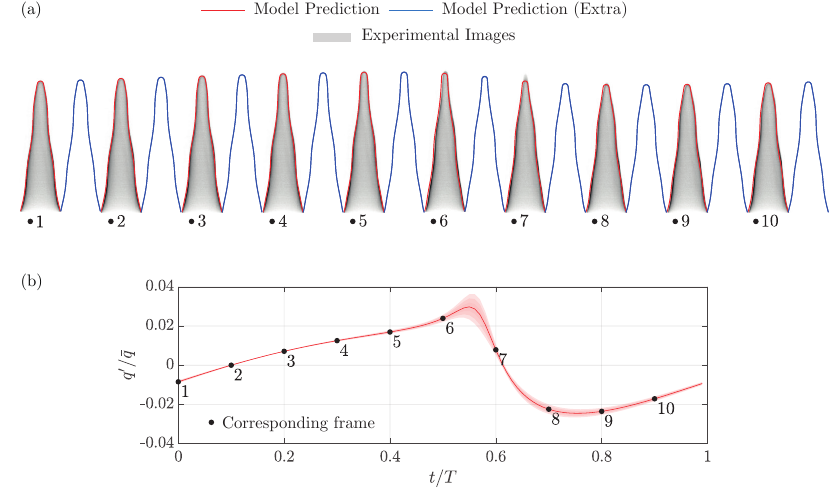}
\end{center}
\caption{Diagram showing the concept of using Bayesian inference to create a quantitatively accurate model capable of providing physical information not directly available from experiments. The experimental data consists of ten frames during one period of oscillation. Panel (a) shows the model prediction (red line) with a confidence interval of 3 standard deviations (red shading) plotted on top of the experimental flame images, as in \ref{fig:UnsteadyPriorAndPosterior}. We also use the model to infer the flame position (blue lines) and its uncertainty (blue shading) at moments between observations. Panel (b) shows the h.r.r. perturbation normalised by its mean value, computed using the model. \CB{The black dots correspond to the timeframes observed in the experiments}}
\label{fig:flameResults}
\end{figure}

Once the most probable parameter values are inferred and the flame front position has been predicted, we can estimate the heat release rate using equation \eqref{eq:compacthrr}. \Yoko{The resulting heat release rate estimation is shown in figure \ref{fig:flameResults}.}
%
Panel (a) shows the model prediction (red lines) with a confidence of 3 standard deviations (red shading) \Yoko{overlaid on the experimental observations (grayscale)}. We also plot the predictions between two observations (blue lines and shading). Panel (b) shows the h.r.r. perturbation $q'$ normalised by its mean value $\bar{q}$, computed using the model. \CB{The black dots correspond to the time frames observed in the experiments}. Figure \ref{fig:flameResults} highlights the advantages of the Bayesian inference approach: using experimental flame images we know the flame front position at ten different time points during a period of oscillation; we use this data to infer the seven parameters of a physics-based reduced order model, which gives the flame front position and the heat release rate at any point in time with their uncertainties. 

\subsection{Step 2: assimilation of all \CB{flames} into a general model}\label{sec:STEP2}
\begin{figure}
\setlength{\fboxsep}{0pt}%
\setlength{\fboxrule}{0pt}%
\begin{center}
\includegraphics[width = 0.75\textwidth]{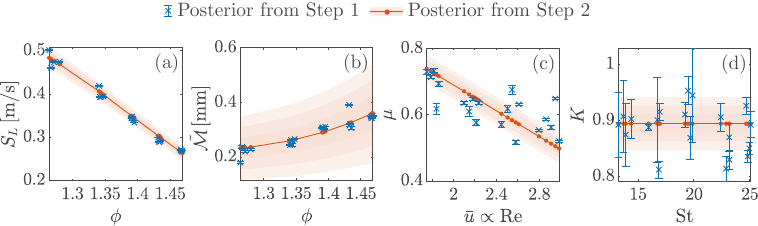}
\end{center}
\caption{Results of the second step of the data assimilation process. From the first step of the process, we obtain the trends of the physical parameters with respect to the equivalence ratio $\phi$ and the bulk velocity $\bar{u}$, which is proportional to the Reynolds number, Re, inside the burner tube. Then we propose a general model for the parameters and assimilate all the flame conditions together.
The blue markers represent the posterior values from the first step of the data assimilation process and the error bars correspond to 3 standard deviations. The red dotted lines correspond to the posterior values from the second step and the shaded areas represent 1, 2 and 3 standard deviations. Panel (a) shows the assimilated values of the unstretched laminar flame speed $S_L$ plotted against the equivalence ratio $\phi$. Panel (b) shows the assimilated values of the dimensional Markstein length $\tilde{\mathcal{M}}$ plotted against the equivalence ratio $\phi$. Panel (c) shows the assimilated values of the velocity shape parameter $\mu$ plotted against the bulk velocity in the burner tube $\bar{u}$. Panel (d) shows the assimilated values of the convective speed ratio $K$ plotted against the acoustic perturbation Strouhal number $\mbox{St}$.
}\label{fig:STEP2_HP}
\end{figure}

\CB{
At this stage, we introduce a linking model $\parameters = \parameters(\phi,\bar{u};\hyperpar)$, designed to determine the model parameters $\parameters$ based on the equivalence ratio $\phi$ and bulk flow rate in the burner tube $\bar{u}$, given any specified set of hyperparameters $\hyperpar$.}
%
Figure \ref{fig:STEP2_HP} shows the \textit{maximum a posteriori} probability distribution of the model parameters $\parameters_\map$ assimilated in the first step, denoted by blue markers with error bars representing 3 standard deviations.
Panel (a) shows the assimilated unstretched laminar flame speed ($S_L$) against the equivalence ratio $\phi$.
Panel (b) shows the assimilated Markstein length ($\mathcal{M}$) against the equivalence ratio $\phi$.
Panel (c) shows the assimilated shape parameter ($\mu$) against the bulk velocity ($\bar{u}$) in the burner tube. The parameter $\mu$ decreases linearly with increasing $\bar{u}$.
Panel (d) shows the assimilated convective speed ratio ($K$) against the acoustic-forcing Strouhal number St. The plot shows that $K$ falls within the range of $0.8$ to $1$, with no discernible pattern with respect to St.
The remaining parameters $\varepsilon_V$, $\lambda$, and $\phi_0$ (not shown) depend on experimental conditions and vary according to external acoustic forcing, both of which are independent of the flame properties.
\par
Based on these observations, the model $\parameters = \parameters(\phi,\bar{u};\hyperpar)$ is formulated as follows:
(i) the laminar flame speed $S_L = S_L^c + \CB{\alpha}_1  \phi + \CB{\alpha}_2$ is equal to the laminar flame speed computed using Cantera $S_L^c$ plus a linear correction that depends on the equivalence ratio; (ii) the Markstein length $\mathcal{M} = \CB{\alpha}_3 \mathcal{M}^c$ is equal to the Markstein length computed using Cantera multiplied by the parameter $\CB{\alpha}_3$; (iii) the shape parameter $\mu = \CB{\alpha}_4 \bar{u} + \CB{\alpha}_5$ is assumed to vary linearly with the bulk flow velocity and therefore with the Reynolds number of the bulk flow in the burner tube; (iv) the value $K = \CB{\alpha}_6$ is assumed to be constant across all flame cases.
We keep $\epsV,\lambda$ and $\phi_0$ equal to the values obtained in the previous assimilation step because the amplitudes of the acoustic forcing and of the flame base oscillations \Yoko{depend on the specific experimental conditions and are not general properties of the flame.}

We then assimilate the most probable hyperparameter set, denoted as $\hyperpar_\map$, by minimizing the cost function expressed by equation \eqref{eq:CostFunc_general} \Yoko{for all flames at once}. The results of this step are shown in red in figure \ref{fig:STEP2_HP}. The red lines display the estimates of the physical parameters $\parameters$ as functions of the equivalence ratio $\phi$, the mean flow velocity $\bar{u}$, and the Strouhal number $\mbox{St}$, after the assimilation of the hyperparameters $\hyperpar$. The three red contours show 1, 2, and 3 standard deviations from the posterior values.

\begin{figure}
\setlength{\fboxsep}{0pt}%
\setlength{\fboxrule}{0pt}%
\begin{center}
\includegraphics[height = 0.65\textwidth]{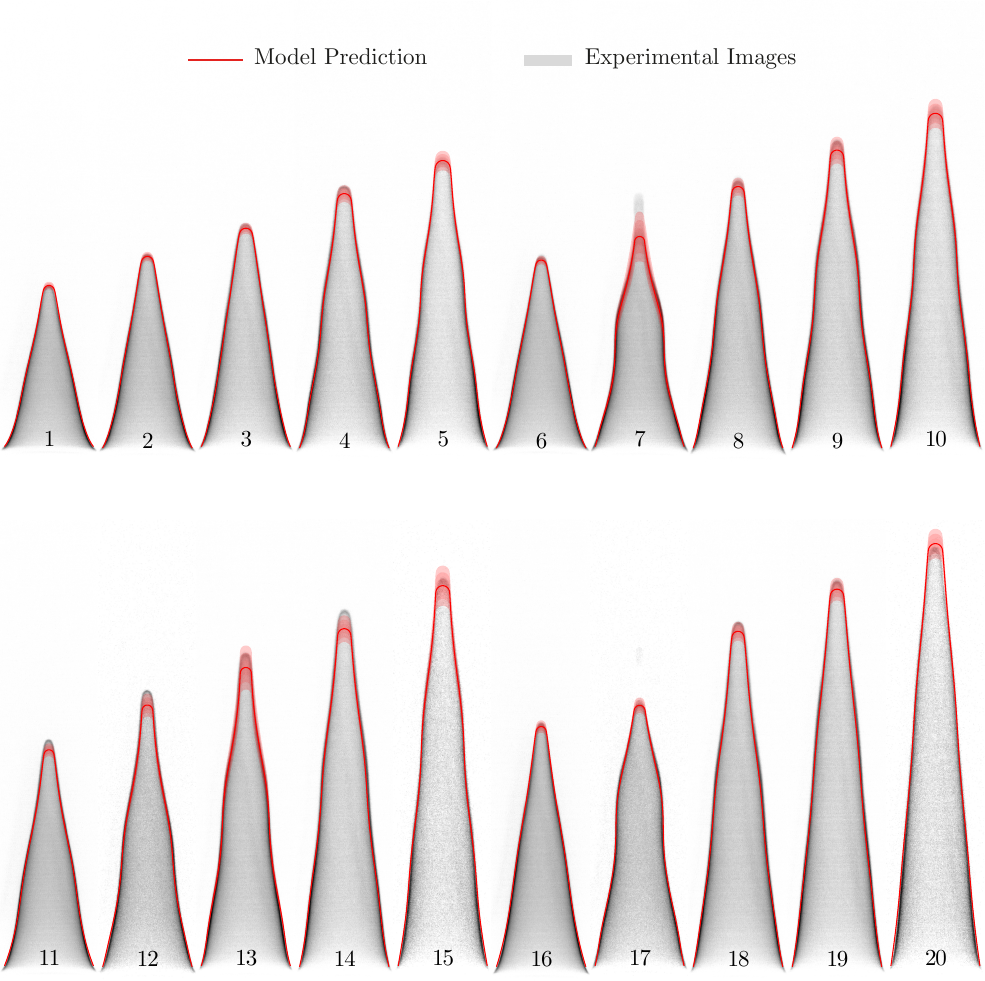}
\end{center}
\caption{
The general model predictions for all flames in the library are represented by the red line, with confidence intervals of 3 standard deviations (depicted as red shading), overlaid on the experimental flame images. The numbers at the base of the flames indicate the corresponding flame numbers referenced in Table \ref{tab:flame_table}.  Additionally, in the supplementary material, we provide an animated version of this figure, showing predictions at all time steps of the experimental video footage.
}\label{fig:STEP2_FLAMES}
\end{figure}

Figure \ref{fig:STEP2_FLAMES} displays the predicted flame front dynamics for all 20 flames in the library at a single timestep. The experimental observations are shown as grayscale images, and the model predictions are overlaid as red lines with confidence intervals of 1, 2, and 3 standard deviations (depicted as red shading). The numbers at the base of the flames indicate the corresponding flame numbers referenced in Table \ref{tab:flame_table}. In the supplementary material, we provide the animated version of this figure, showing the predictions for all timesteps of the experimental video footage.

\subsection{Flame transfer function predictions}\label{sec:FTF}
\CB{We introduce the flame transfer function $\mathcal{F}$ as:
\begin{equation}
\mathcal{F}(\mbox{St};\phi, \bar{u},\hyperpar) = \frac{\hat{q}/\bar{q}}{\hat{u}_{ref}/\bar{u}},
\end{equation}
where $\hat{\star}$ denotes the Fourier component of $\star$ at the Strouhal number $\mbox{St}$; $q$ denotes the fluctuating heat release rate and $u_{ref}$ denotes the velocity fluctuations at a reference position. In this study, we define the reference velocity $u_{ref}$ as the component of the velocity field perpendicular to the flame front at its base, adhering to the definition provided by Orchini \& Juniper \citep{Orchini2016a}. The heat release rate $q$ is normalised by the mean heat release rate, while the velocity $u$ is normalised by the mean flow velocity in the burner tube $\bar{u}$.}
We use the calibrated general model obtained in the second step to predict the gain and phase of the flame transfer function at the excitation Strouhal number $\mbox{St}$ for all flames in \CB{Table \ref{tab:flame_table}}.

\begin{figure}
\setlength{\fboxsep}{0pt}%
\setlength{\fboxrule}{0pt}%
\begin{center}
\includegraphics[width = .7\textwidth]{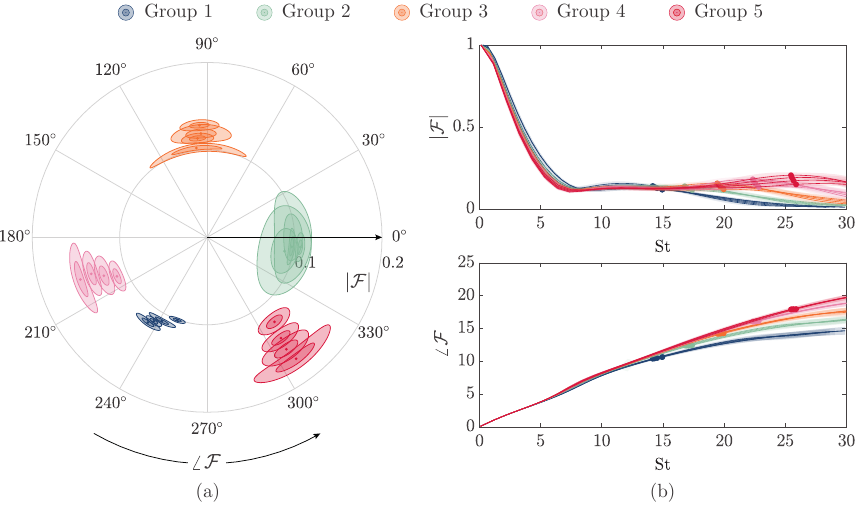}
\end{center}
\caption{ Inferred flame transfer functions presented as (a) polar plots, which show gain on the radial axis and phase on the angular axis, and (b) Bode plots, which show (i) gain and (ii) phase against Strouhal number St. In (a) we plot flame transfer functions as shaded patches, which denote 1-3 standard deviations around the expected value. In (b) we plot the flame transfer functions as solid lines with a shaded patch representing the confidence bounds of 3 standard deviations, whereas the coloured dots represent the values corresponding to the experimental forcing frequency.}\label{fig:FTFs_1}
\end{figure}

In figure \ref{fig:FTFs_1}(a), we present the model predictions along with their uncertainties in a polar plot. Each flame transfer function is coloured according to its corresponding flame group \CB{as in figure \ref{fig:flameLibraryIllustration}}. It is worth noting that the flames within the same group, which are grouped based on \Yoko{convective} time delay $\tau$, exhibit similar phase delays on the polar plot, \CB{as one would expect}. 
We extend these findings by assuming that the hyperparameters $\hyperpar$ remain constant across different forcing frequencies. Figure \ref{fig:FTFs_1}(b) illustrates the predictions of flame transfer functions along with their uncertainties over a wide range of Strouhal numbers ($\mbox{St}$). In future research, we will evaluate the validity of this assumption \Yoko{by assimilating data collected over a range of Strouhal numbers}. Should it prove untenable, we will enhance the complexity of the linking model $\parameters = \parameters(\phi,\bar{u};\hyperpar)$ to incorporate the impact of the forcing frequency, by following the same methodology presented in this paper.

%
%
%

\subsection{Extrapolation}\label{sec:Extrapolation}
\begin{figure}
\setlength{\fboxsep}{0pt}%
\setlength{\fboxrule}{0pt}%
\begin{center}
\includegraphics[width = 0.65\textwidth]{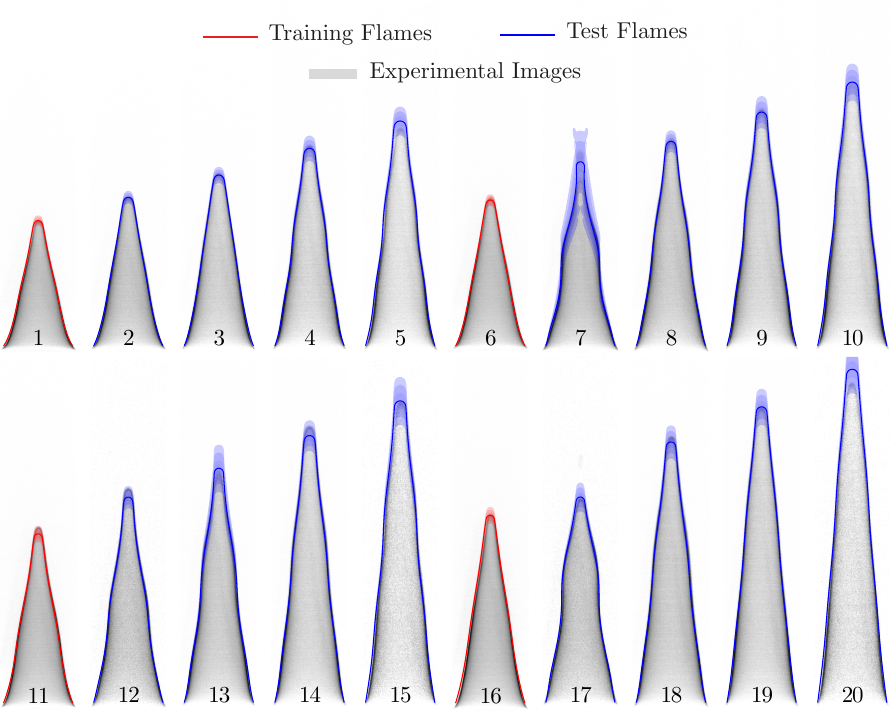}
\end{center}
\caption{Predicted flame front dynamics for all flames in the library at a single timestep. Red lines show flames used for training, while blue lines show flames used for testing. The shaded regions in red and blue indicate 1, 2, and 3 standard deviations from the optimal flame front estimation. Additionally, all model predictions are overlaid onto the experimental flame images for comparison. In the supplementary material, we provide an animated version of this figure, presenting predictions for all time steps of the experimental video footage.
}\label{fig:STEP3_FLAMES}
\end{figure}

The model in this paper contains a small number of parameters because it is physics-based. 
Consequently, the training process requires only a small amount of data. 
In this section, we show the model's ability to predict the flame front dynamics of the 20 flames in our library (see Table \ref{tab:flame_table}) using data from just four flames. 
We train the model's hyperparameters on flames 1, 6, 11, and 16 of Table \ref{tab:flame_table} and predict the remaining flames with this model. \Yoko{These flames have the same convective time delay, but different powers. We therefore test the ability of the model to extrapolate outside the range of trained convective time delays.}
Figure \ref{fig:STEP3_FLAMES} shows the predicted flame front dynamics for all flames at a single timestep. Red lines show the flames used for training. Blue lines show the flames used for testing. Red and blue shading denotes 1, 2, and 3 standard deviations from the \Yoko{predicted flame front position}. All model predictions are overlaid onto the experimental flame images for comparison.
The number at the base of each flames indicates the flame numbers in Table \ref{tab:flame_table}. In the supplementary material, we provide the animated version of this figure, showing the predictions for all timesteps of the experimental video footage.
\CB{We now compare figure \ref{fig:STEP3_FLAMES} with figure \ref{fig:STEP2_FLAMES}. When only one-fifth of the available data has been used, the predicted flame front positions are less accurate and have larger uncertainty. This is particularly clear for flame 7, which has a large acoustic amplitude and is close to pinch off in these figures. Indeed, the model shows significant sensitivity to the model parameters as the acoustic amplitude increases. Similarly, the prediction for flame 20 is less accurate because the flame length is much larger than the length of the four flames used for model training. Even with such a small amount of training data, however, the data still falls within the uncertainty bounds of the model predictions, even when the data is well outside the training range. We further note that figure \ref{fig:STEP3_FLAMES} shows the timestep with the largest model misfit and uncertainties. In the supplementary material, we provide an animated version of this figure showing the predictions for all timesteps of the video footage.
}

\section{Conclusions}\label{sec:Conclusions}
In this study, we perform experiments on an acoustically forced, laminar premixed conical flame in a duct. We use a high-speed camera to record snapshots of the natural emission of the flame while steady and forced. We propose a physics-based reduced-order model that provides, amongst other things, the flame front dynamics under harmonic forcing. We combine the model output with the experimental images of the flame to infer the most probable model parameters. This process (i) turns a qualitatively-accurate model into a quantitatively-accurate model, and (ii) quantifies the uncertainty in the inferred model parameters and the model predictions.

The inference process produces a digital twin of the flame, which provides access to quantities that were not directly measured in the experiments. From observations of the steady flame, we can estimate combustion characteristics such as the laminar flame speed and the Markstein length. This can be used to characterise the combustion properties of a fuel for which data are not available. 
From observations of the perturbed flame, we can infer the fluctuating heat release rate as a response to the velocity perturbation, which is then used to calculate the flame's thermoacoustic response through, for example, the flame transfer function.
The proposed model contains few parameters and can therefore be trained on relatively little data. To demonstrate this we infer the model's parameters using one-fifth of the available data and successfully predict the remaining flames in the dataset. 
\CB{The model is also differentiable with respect to the parameters meaning that it can be used as a design optimization tool.}
%

In future work, the flame model will be included within a larger model
of the thermoacoustic behaviour of the system, whose other parameters are
tuned in the same way. The final result will be a quantitatively-accurate physics-based model of the flame and the thermoacoustic system that is interpretable, trustworthy, and extrapolatable.

The Matlab code developed in this study is made available to the reader, who is invited to modify and apply it to assimilate other flame image data into the flame model.

\section*{Declaration of competing interest}
The authors declare that they have no known competing financial interests or personal relationships that could have appeared to influence the research reported in this paper.

\section*{Data availability}
The experimental data is  openly available at http://doi.org/10.17863/CAM.106960. The Matlab code is available at \url{https://github.com/aleGiannotta/Bayesian_Inference_Lam_Flame.git}

\section*{Acknowledgements}
This research was partly supported under the National Recovery and Resilience Plan (NRRP), Mission 4 Component 2 Investment 1.4 - Call for “National Centres” from research to business, funded by the European Union – NextGenerationEU. Project code CN00000023, Concession Decree No. 1033 adopted by Ministero dell’Università e della Ricerca (MUR), CUP - D93C22000410001, Project title ‘‘MOST - National Center for Sustainable Mobility’’. 
This research was partly
carried out within the NEST - Network 4 Energy Sustainable Transition (D.D. 1243 02/08/2022, PE00000021) and received funding under the National Recovery and Resilience Plan (NRRP), Mission 4 Component 2 Investment 1.3, funded from the European Union - NextGenerationEU. This manuscript reflects only the authors’ views and opinions, neither the European Union nor the European Commission can be considered responsible for them.
Matthew Yoko acknowledges funding for his PhD from The Cambridge Trust, The Skye Foundation and the Oppenheimer Memorial Trust.
\appendix
\section{Properties of the flames studied in this paper}
\label{sec:app_a}
\begin{minipage}{\textwidth}
\centering 
\captionof{table}{Summary of the properties of the 20 flames studied. We show the average measured flow rates of air, methane ($\mathrm{CH_4}$) and ethylene ($\mathrm{C_2H_4}$), the equivalence ratio ($\phi$), the bulk velocity in the burner tube ($\bar{u}$), the measured flame length ($L_f$), and the mean heat release rate ($\bar{Q}$).}
\label{tab:flame_table} 
\begin{tabular*}{\textwidth}{l@{\extracolsep{\fill}}lllllllll} 
\hline \\ [-1em] 
Flame No. & Air      & $\mathrm{CH_4}$ & $\mathrm{C_2H_4}$ & $\phi$ & $\bar{u}$ &  $L_f$ meas.  & $\bar{Q}$  \\
{[}-]     & [ln/min] & [ln/min]         & [ln/min]           & [-]     & [m/s]      & [mm]          & [W] \\  \\[-1em] 
\hline 
1 &  6.049 & 0.325 & 0.325 & 1.28 & 1.75 & 16.6 & 375 \\  
2 &  6.147 & 0.348 & 0.348 & 1.35 & 1.79 & 20.0 & 375 \\  
3 &  6.219 & 0.364 & 0.364 & 1.40 & 1.82 & 23.1 & 375 \\  
4 &  6.283 & 0.379 & 0.379 & 1.44 & 1.84 & 26.4 & 375 \\  
5 &  6.338 & 0.391 & 0.391 & 1.47 & 1.86 & 28.9 & 375 \\  
\hline
6 &  7.246 & 0.387 & 0.387 & 1.27 & 2.10 & 19.8 & 450 \\
7 &  7.369 & 0.416 & 0.416 & 1.34 & 2.15 & 23.8 & 450 \\
8 &  7.459 & 0.436 & 0.436 & 1.39 & 2.18 & 27.4 & 450 \\
9 &  7.537 & 0.454 & 0.454 & 1.43 & 2.21 & 30.9 & 450 \\
10 &  7.603 & 0.468 & 0.468 & 1.47 & 2.24 & 34.2 & 450 \\
\hline
11 &  8.444 & 0.449 & 0.449 & 1.27 & 2.45 & 23.1 & 525 \\
12 &  8.594 & 0.484 & 0.484 & 1.34 & 2.51 & 27.6 & 525 \\
13 &  8.699 & 0.508 & 0.508 & 1.39 & 2.55 & 31.7 & 525 \\
14 &  8.790 & 0.529 & 0.529 & 1.43 & 2.58 & 36.4 & 525 \\
15 &  8.868 & 0.546 & 0.546 & 1.47 & 2.61 & 39.2 & 525 \\
\hline
16 &  9.644 & 0.512 & 0.512 & 1.26 & 2.80 & 25.9 & 600 \\
17 &  9.818 & 0.553 & 0.553 & 1.34 & 2.86 & 30.1 & 600 \\
18 &  9.939 & 0.580 & 0.580 & 1.39 & 2.91 & 34.8 & 600 \\
19 &  10.045 & 0.604 & 0.604 & 1.43 & 2.95 & 39.9 & 600 \\
20 &  10.134 & 0.624 & 0.624 & 1.47 & 2.99 & 43.6 & 600 \\
\hline 
\end{tabular*} 
\end{minipage}

 \bibliographystyle{elsarticle-num}
 \bibliography{references}





\end{document}